\documentclass{article}
\pdfoutput=1

\usepackage{jheppub}

\usepackage{adjustbox}
\usepackage{amsmath}
\usepackage{booktabs}
\usepackage{braket}
\usepackage{colortbl} 
\usepackage{comment}
\usepackage{dcolumn}
\usepackage{graphicx}
\usepackage{hyperref}
\usepackage[utf8]{inputenc}
\usepackage{multirow}
\usepackage{siunitx}
\usepackage[normalem]{ulem}
\usepackage{xcolor}
\usepackage{xspace}
\usepackage{pdflscape}

\newcommand{\gev}{\textrm{GeV}\xspace}
\newcommand{\mev}{\textrm{MeV}\xspace}
\newcommand{\bsk}{\ensuremath{\bar{B}_s\rightarrow K}\xspace}
\newcommand{\lhcbdecay}{\ensuremath{B_s^0\rightarrow K^-\mu^+\nu_\mu}\xspace}
\newcommand{\dynesty}{\texttt{dynesty}\xspace}
\newcommand{\EOS}{\texttt{EOS}\xspace}
\newcommand{\Vcb}{\ensuremath{V_{cb}}\xspace}
\newcommand{\absVcb}{\ensuremath{\left| \Vcb \right| }\xspace}
\newcommand{\Vub}{\ensuremath{V_{ub}}\xspace}
\newcommand{\absVub}{\ensuremath{\left| \Vub \right| }\xspace}

\makeatletter
\let\oldtheequation\theequation
\renewcommand\tagform@[1]{\maketag@@@{\ignorespaces#1\unskip\@@italiccorr}}
\renewcommand\theequation{(\oldtheequation)}
\makeatother

\begin{document}

\title{New determination of $|V_{ub}/V_{cb}|$ from $B_s^0\to \lbrace K^-, D_s^- \rbrace \mu^+\nu$}
\author[a,b]{Carolina Bolognani,}
\emailAdd{carolina.bolognani@cern.ch}
\author[c]{Danny van Dyk,}
\emailAdd{danny.van.dyk@gmail.com}
\author[a,b]{K. Keri Vos}
\emailAdd{k.vos@maastrichtuniversity.nl}

\affiliation[a]{Gravitational Waves and Fundamental Physics (GWFP), Maastricht University, Duboisdomein 30, NL-6229 GT Maastricht, the Netherlands}
\affiliation[b]{Nikhef, Science Park 105, NL-1098 XG Amsterdam, the Netherlands}
\affiliation[c]{Institute for Particle Physics Phenomenology and Department of Physics, Durham University, Durham DH1 3LE, UK}

\abstract{
We update the full set of \bsk form factors using light-cone sum rules with an on-shell kaon.
Our approach determines the relevant sum rule parameters---the duality thresholds---from a Bayesian fit for the first time.
Using a modified version of the Boyd-Grinstein-Lebed parametrisation, we combine our sum rule results at low momentum transfer $q^2$
with more precise lattice QCD results at large $q^2$.
We obtain a consistent description of the form factors in the full $q^2$ range.
Applying these results to a recent LHCb measurement of branching ratios for the decays $B_s^0 \to \lbrace K^-, D_s^-\rbrace \mu^+\nu_\mu$,
we determine the ratio of Cabibbo-Kobayashi-Maskawa elements
$$
\notag
\left|\frac{V_{ub}}{V_{cb}}\right|_{q^2<7\; \gev^2} = 0.0681\pm 0.0040
\quad \text{and} \quad
\left|\frac{V_{ub}}{V_{cb}}\right|_{q^2>7 \;\gev^2} = 0.0801\pm 0.0047 \ ,
$$
which are mutually compatible at the $1.9\sigma$ level.
We further comment on the sensitivity to Beyond the Standard Model effects through measurements of the shape
of $B_s^0 \to K^- \mu^+\nu_\mu$ decays, in light of recent limits on such effects from other exclusive $b\to u\ell\nu$ processes.
}

\begin{flushright}
    EOS-2023-03\\
    IPPP/23/42\\
    Nikhef-2023-08\\
\end{flushright}
\vspace*{-3\baselineskip}

\maketitle

\section{Introduction}
\label{sec:intro}

Quark flavour mixing, as described by the Cabibbo-Kobayashi-Maskawa (CKM) quark mixing matrix, is a central paradigm
of the Standard Model (SM) of particle physics. The CKM matrix elements are not predicted by the SM and require
determination from experimental data, which uses hadronic matrix elements as essential theoretical inputs.
The determination of the CKM elements $V_{cb}$ and $V_{ub}$ is beset by ongoing puzzles, since determinations
of these quantities from inclusive and exclusive $B$-meson decays differ systematically;
see e.g. \cite{Gambino:2020jvv} for a recent review and \cite{Gambino:2019sif,Bordone:2021oof,Belle:2021eni,Leljak:2021vte,Bernlochner:2022ucr,Biswas:2022yvh,Leljak:2023gna,Greljo:2023bab} for recent determinations. 
The ongoing puzzles reflect
both the experimental difficulties in measuring these decays and the theoretical difficulties in providing the
essential hadronic information to extract the matrix elements.\\

Recently, the LHCb collaboration measured 
$\mathcal{B}(B_s^0\to K^- \mu^+ \nu_\mu)/\mathcal{B}(B_s^0 \to D_s^-\mu^+\nu_\mu)$~\cite{LHCb:2020ist}
and then extracted the ratio of CKM elements $\absVub/\absVcb$, using specific hadronic inputs for the $\bsk$ transition.
Recent theory developments~\cite{Flynn:2023nhi,Flynn:2023qmi} call for an update of this determination.
We aim to study the effects of all the available theory information on this extraction by updating the hadronic $\bsk$ form factors,
which enter all theoretical predictions of the ratio of branching fractions.\\

The full set of \bsk form factors at mass dimension three is defined as
\begin{align}
    \label{eq:intro:ff-def}
    \bra{K^+(k)} \bar{u} \gamma^\mu b \ket{\bar{B}_s(p)}
        & = f_+(q^2) \left[(p+k)^\mu - \frac{m_{B_s}^2 - m_K^2}{q^2}q^\mu\right] + f_0(q^2) \frac{m_{B_s}^2-m_K^2}{q^2}q^\mu \\
    \bra{K^+(k)} \bar{u} \sigma_{\mu\nu} b \ket{\bar{B}_s(p)} & = \frac{if_T(q^2)}{m_{B_s} +m_K} \left[q^2(p+k)_\mu - (m_{B_s}^2-m_K^2)q_\mu\right] \ ,
\end{align}
where $q^\mu \equiv p^\mu - k^\mu$ denotes the momentum transfer to the lepton-neutrino pair.
A kinematic singularity in the matrix element of the vector current is avoided by the identity
$f_+(q^2=0)=f_0(q^2=0)$.
For the extraction of CKM matrix elements, only the form factors $f_+$ and $f_0$ are needed, with
$f_0$ taking a numerically subleading role if the charged lepton final state $\ell$ is light, i.e., $\ell=e,\mu$.
However, to probe for effects Beyond the Standard Model (BSM), the
form factor $f_T$ becomes relevant. Moreover, the \bsk form factors are related by isospin symmetry to the \bsk form factors entering rare neutral-current $b\to d \ell^+\ell^-$ processes.
In the latter case, $f_T$ is essential for SM predictions of the decay.
Hence, we include the form factor $f_T$ in our analysis.

For \bsk form factors there is some disagreement in the literature, both between individual determinations
from different Lattice QCD collaborations, and  between Lattice QCD and light-cone sum rule (LCSR) analyses.
In particular, for small values of the momentum transfer $q^2$, the situation can be summarized as follows:
\begin{itemize}
  \item LCSR analyses of these form factors work best at small to negative values of $q^2$.
    Hence, their predictions of $f_+(q^2 = 0)$ can be extracted directly and do not require extrapolation.
    A 2008 analysis~\cite{Duplancic:2008tk} yields
    \begin{equation}
        \label{eq:intro:fp-DM2008}
        f_+(q^2 = 0)|_\text{DM2008} = 0.30^{+0.04}_{-0.03}\ ,
    \end{equation}
    which has since been superseded by an updated analysis~\cite{Khodjamirian:2017fxg}
    \begin{equation}
        \label{eq:intro:fp-KR2017}
        f_+(q^2 = 0)|_\text{KR2017} = 0.336 \pm 0.023\ .
    \end{equation}
  \item In the HPQCD Lattice QCD analysis~\cite{Bouchard:2014ypa}, the extrapolation to $q^2 = 0$ yields
    \begin{equation}
        f_+(q^2 = 0)|_\text{HPQCD2014} = 0.323 \pm 0.063\ .
    \end{equation}
    \item In both the FNAL/MILC analysis~\cite{FermilabLattice:2019ikx} and a (by now superceded) RBC/UKQCD analysis~\cite{Flynn:2015mha},
    the extrapolation to $q^2 = 0$ yields very small values at $q^2$:
    \begin{equation}
    \begin{aligned}
        f_+(q^2 = 0)|_\text{FNAL/MILC2019} & = 0.13  \pm 0.05      \\
        f_+(q^2 = 0)|_\text{RBC/UKQCD2015} & = 0.159 \pm 0.059 \ .
    \end{aligned}
    \end{equation}
  \item Very recently, an updated RBC/UKQCD analysis~\cite{Flynn:2023nhi} has been published that uses a different approach
    for the chiral and continuum extrapolation of the form factors, changing from the procedure
    also used in Ref.~\cite{FermilabLattice:2019ikx} and adopting a similar procedure as used in Ref.~\cite{Bouchard:2014ypa}\footnote{%
        Both Ref.~\cite{Bouchard:2014ypa} and Ref.~\cite{Flynn:2023nhi}, use a basis consisting of the $f_+$ and $f_0$ form factors, but the two works differ
        in the extrapolation to the continuum limit.
    }.
    The new RBC/UKQCD analysis also uses a different form factor parametrisation for the extrapolation to small values of $q^2$~\cite{Flynn:2023qmi},
    based on earlier works on dispersive form factor bounds in presence of sub-threshold
    branch points~\cite{Gubernari:2020eft,Blake:2022vfl,Amhis:2022vcd}.
    This work yields
    \begin{equation}
        f_+(q^2 = 0)|_\text{RBC/UKQCD} = 0.25 \pm 0.11 \ .
    \end{equation}
\end{itemize}

In light of these discrepancies, it is not surprising that LHCb finds mutually incompatible results for the ratio $|V_{ub}|/|V_{cb}|$
in the two bins of $q^2$ that are analysed. Concretely, LHCb finds~\cite{LHCb:2020ist}
\begin{align}\label{eq:lhcbrat}
    |V_{cb}/V_{ub}|_{q^2<7 \;{\rm GeV}^2}
        & =  0.061 \pm 0.004 \ , \\
    |V_{cb}/V_{ub}|_{q^2>7 \;{\rm GeV}^2}
        & =  0.095 \pm 0.008 \ ,
\end{align}
in the two available $q^2$ bins. These results are based on the aforementioned lattice QCD inputs by FNAL/MILC~\cite{FermilabLattice:2019ikx} at large $q^2$ and the LCSR inputs by
KR2017~\cite{Khodjamirian:2017fxg} at low $q^2$ for the \bsk form factors,
and the lattice QCD inputs by HPQCD~\cite{McLean:2019qcx} for $\bar{B}_s\to D_s$ in its entire kinematic region.
Since these determinations are dominated by the form factor input, the difference between the two calls for a close examination of the theoretical inputs. 
The purpose of this article is to revisit the LCSR analysis of the \bsk form factors and to perform a global fit to the available
form factor information along the lines of a previous analysis of $\bar{B}\to \pi$ form factors~\cite{Leljak:2021vte} and
to clarify the situation in the \bsk form factors.
Subsequently, using the existing lattice QCD determination of the $\bar{B}_s\to D_s$ form factors by the HPQCD collaboration~\cite{McLean:2019qcx} and the measured LHCb data,
we update the determination of the ratio $|V_{ub}|/|V_{cb}|$.

\section{The \bsk form factors from light-cone sum rules}
\label{sec:lcsr}

We determine the \bsk form factors using Light-Cone Sum Rule~(LCSRs)~\cite{Braun:1988qv,Balitsky:1989ry,Chernyak:1990ag, Colangelo:2000dp} techniques. The LCSR is set up by defining a tailored two-point correlation function, e.g.,
\begin{equation}
    \label{eq:lcsr:correlator}
    i\int d^4x e^{iqx} \bra{K(k)} T\lbrace J_{B_s}(x), [\bar{u} \gamma^\mu b](0)\rbrace \ket{0}
        = \sum_{t,n} \int \mathcal{D}u \, T_n(k, q, \vec{u}) \;\phi_{t,n}(\vec{u})\,.
\end{equation}
This correlation function factorizes into perturbative~(hard) scattering kernels $T_n$ and universal nonperturbative
light-cone distribution amplitudes~(LCDAs) $\phi_{t,n}$
if the integral on the left-hand-side in \autoref{eq:lcsr:correlator} is dominated by light-like distances $x^2 \simeq 0$.
The integral on the right-hand side involves the fractions of the kaon momentum
carried by the partons, i.e., of the quark and antiquark in the two-particle Fock state, and the quark, antiquark and gluon in the three-particle Fock state.
The integration measure reads
\begin{equation}
    \int \mathcal{D}u = \int \delta(1 - \sum_i u_i) \prod_i du_i\,.
\end{equation}

The factorisation is achieved by means of a light-cone operator product expansion (LCOPE).
The LCSR is then constructed by connecting the correlation function in \autoref{eq:lcsr:correlator}
with one or more of the hadronic form factors in \autoref{eq:intro:ff-def} using a dispersion relation and assuming semi-global quark-hadron duality.
Here, we construct the LCSRs using an on-shell kaon state and interpolating the $\bar{B}_s$ meson
with an interpolating current $J_{B_s}$. As a consequence, our setup relies on the LCDAs of the kaon~\cite{Ball:2006wn,LatticeParton:2022zqc}.

The power counting within the LCOPE is achieved in terms of the operators' twist $t$.
This is different than in a local operator product expansion, where the operators' mass dimension is the relevant quantity.
Within the LCOPE, contributions due to operators with twist $t$ are suppressed by powers of
$\left(\Lambda_\text{had}/E\right)^{t - 2}$ with respect to the leading twist-2 terms \cite{Braun:1999uj}, with $E$ a large
energy scale associated with the momentum transfer $q^2$ of the form factors.
Beyond the two-particle level, different contributions of the same twist are enumerated by the index $n$.
The LCSRs for \bsk form factors are known to high accuracy: two-particle twist-2 and twist-3 terms are known
to next-to-leading order in $\alpha_s$, and two-particle twist-4 terms are known to leading order;
three-particle terms at twists 3 and 4 are also known to leading order~\cite{Duplancic:2008tk}.
Contributions at the twist-5 and twist-6 level are estimated in Ref.~\cite{Khodjamirian:2017fxg} using a factorisation
approximation~\cite{Rusov:2017chr} and found to be negligible. Therefore, we do not include these terms
in our analysis.\\

We implement the LCSRs for the full basis of form factors.
Our implementation is independent of any specific choice of model for the LCDAs by using the parametrisations
provided in Ref.~\cite{Ball:2006wn}.
As discussed in Ref.~\cite{Duplancic:2008tk}, the effects of a non-zero strange quark and kaon mass become
relevant in the sum rule analysis. The results for \bsk form factors can be inferred from
the well-known results for the $\bar{B}\to \bar{K}$ LCSR by
\begin{itemize}
    \item exchanging the strange quark mass with the spectator quark mass, $m_s \leftrightarrow m_q$
    \item exchanging the quark and antiquark momentum quantities, i.e. $u \leftrightarrow \bar{u} = 1 - u$
    for two-particle LCDAs, and $\alpha_1 \leftrightarrow \alpha_2$ for three-particle LCDAs.
\end{itemize}
Using the parametrisations and renormalisation group equations provided in Ref.~\cite{Ball:2006wn},
we perform a trivial cross check by applying the above exchanges twice: once analytically in our numerical code,
and once numerically by exchanging the values of the quark masses and changing the numerical values of the kaon LCDA parameters.
The latter only involves changing the sign of the odd LCDA coefficients, e.g., $a_{2n + 1K}$ and similar.
We initially find that our numerical code does not fulfill this cross check for all values of the renormalisation scale,
except for the nominal scale of $\mu = 1\,\gev$. We identify the terms proportional to the strange quark mass
in Eq. (3.11) of Ref.~\cite{Ball:2006wn} as the origin of the problem, since they are expanded to leading order
in $m_q / m_s$. To restore the correct behaviour under the cross check, we use that $m_s$ enters the RGE only in the
combinations of $m_s \pm m_q$, which are even (odd) under the exchange of quark and antiquark inside the kaon.
Using the known parity of the Gegenbauer coefficients, we apply the replacements
\begin{equation}
\begin{aligned}
    f_K m_s        & \!\to f_K (m_s + m_q)\ ,     &
    f_K m_s a_{1K} & \!\to f_K (m_s - m_q) a_{1K}\ , &
    f_K m_s a_{2K} & \!\to f_K (m_s + m_q) a_{2K}\ ,
\end{aligned}
\end{equation}
which leads our numerical implementation to pass the aforementioned cross check.
\\

As a central part of our work we update the predictions for the three hadronic form factors
defined in \autoref{eq:intro:ff-def}.
Our numerical results differ from previous LCSR determinations \cite{Duplancic:2008tk,Khodjamirian:2017fxg} due to
updated input parameters as discussed in \autoref{sec:LCSR-inputs} and our determination of the duality thresholds as discussed in \autoref{sec:LCSR-duality}.

\subsection{Input parameters}
\label{sec:LCSR-inputs}

\begin{table}[t!]
\centering
\begin{tabular}{c c c c c}
  \toprule
  parameter                   & value/interval      & unit        & prior       & comments/source \\
  \midrule
  \multicolumn{5}{c}{strong coupling and quark masses}\\
  \midrule
  $\alpha_s \left(m_Z\right)$ & $0.1179 \pm 0.0009$ & \textemdash & gaussian    & \cite{ParticleDataGroup:2022pth} \\
  $\bar{m}_b\left(\bar{m}_b\right)$
                              & $4.18\pm0.03$       & \gev        & gaussian    & \cite{ParticleDataGroup:2022pth} \\
  $m_s\left(2\,\gev\right)$   & $93.4\pm 8.6$       & \mev        & gaussian    & \cite{ParticleDataGroup:2022pth} \\
  $m_u\left(2\,\gev\right)$   & $2$                 & \mev        & \textemdash & fixed    \\
  \midrule
  \multicolumn{5}{c}{hadronic parameters of the $B_s$ and $K$ mesons}\\
  \midrule
  $f_{B_s}$                   & $230.3 \pm 1.3$     & \mev        & \textemdash & \cite{FlavourLatticeAveragingGroupFLAG:2021npn} \\
  $f_K$                       & $155.7 \pm 0.3$     & \mev        & gaussian    & \cite{FlavourLatticeAveragingGroupFLAG:2021npn} \\
  $a_{1K}(1\,\gev)$             & $-0.130 \pm 0.06$   & \textemdash & gaussian    & \cite{LatticeParton:2022zqc} \\
  $a_{2K}(1\,\gev)$             & $0.228 \pm 0.07$    & \textemdash & gaussian    & \cite{LatticeParton:2022zqc} \\
  $f_{3K}(1\,\gev)$             & $[0.003, 0.006]$    & $\gev^2$    & uniform     & \cite{Ball:2006wn} \\
  $\omega_{3K}(1\,\gev)$        & $[-1.9,-0.5]$       & \textemdash & uniform     & \cite{Ball:2006wn} \\
  $\lambda_{3K}(1\,\gev)$       & $[1.2,2.0]$         & \textemdash & uniform     & \cite{Ball:2006wn} \\
  $\delta^2_{K}(1\,\gev)$       & $[0.14,0.26]$       & $\gev^2$    & uniform     & \cite{Ball:2006wn} \\
  $\kappa_{4K}(1\,\gev)$        & $[-0.11,-0.07]$     & \textemdash & uniform     & \cite{Ball:2006wn} \\
  $\omega_{4K}(1\,\gev)$        & $[0.1,0.3]$         & \textemdash & uniform     & \cite{Ball:2006wn} \\
  \midrule
  \multicolumn{5}{c}{sum rule parameters and scales}\\
  \midrule
  $\mu$                       & $3.0$               & \gev        & \textemdash & \\
  $M^2$                       & $[13.0,21.0]$       & $\gev^2$    & uniform     & \cite{Khodjamirian:2017fxg} \\
  $s_0^{f_+}$                 & $[34.5,46.5]$       & $\gev^2$    & uniform     & \\
  $s_0^{f_0}$                 & $[34.5,46.5]$       & $\gev^2$    & uniform     & \\
  $s_0^{f_T}$                 & $[34.5,46.5]$       & $\gev^2$    & uniform     & \\
  $s_0^{'\, f_+}$             & $[-1.0,+1.0]$       & \textemdash & uniform     & \\
  $s_0^{'\, f_0}$             & $[-1.0,+1.0]$       & \textemdash & uniform     & \\
  $s_0^{'\, f_T}$             & $[-1.0,+1.0]$       & \textemdash & uniform     & \\
  \bottomrule
\end{tabular}
\caption{%
  Input parameters used in the numerical analysis of the LCSRs for \bsk form factors.
  We quote individual components of the full prior probability density, which is an uncorrelated product of individual uniform or Gaussian components. Gaussian components cover the stated interval at $68\%$ probability
  For practical purpose, variates of the gaussian priors are only sampled inside their respective central $99\%$ probability intervals.
}
\label{tab:inputs}
\end{table}

Our setup follows the Bayesian approach proposed in Ref.~\cite{SentitemsuImsong:2014plu} to calculate the full set of \bsk form factors
in LCSR. We construct a prior probability distribution for all relevant input parameters, and a theoretical
likelihood for the determination of the duality thresholds. Contrary to Ref.~\cite{SentitemsuImsong:2014plu,Leljak:2021vte},
we do not determine the initial state's decay constant from a two-point sum rule. Instead, we use the world average
of lattice QCD results for the $B_s$ decay constant for $N_f=2+1+1$ flavours~\cite{FlavourLatticeAveragingGroupFLAG:2021npn}.
We classify the full set of input parameters as follows:
 \begin{description}
     \item[\textbf{strong coupling and quark masses}] These include the strong coupling at an initial scale $\mu=M_Z$, bottom quark mass in the $\overline{\text{MS}}$ scheme at the scale $m_b$, the strange quark mass and the sum of up and down quark masses in the $\overline{\text{MS}}$ scheme at the scale $2 \,\gev$.
          
     \item[\textbf{parameters of the $\boldsymbol{K}$ LCDAs}] These include the kaon decay constant $f_K$, which is
     used to normalize the leading-twist LCDA whose shape is described by a Gegenbauer polynomial expansion.
     We keep only the first two terms of the Gegenbauer expansion, and vary their coefficients $a_{1K}$ and $a_{2K}$
     as a Gaussian prior based on information extracted from Ref.~\cite{LatticeParton:2022zqc}.
     We evolve these from the renormalisation scale of $2\,\gev$ to our reference scale of $1\,\gev$ to leading-logarithmic accuracy.
     Following Ref.~\cite{Ball:2006wn}, we normalize the twist-3 two-particle LCDAs with the chiral parameter
     $\mu_K(\mu) = m_K^2 / [m_s(\mu) + m_q(\mu)]$,
     and the twist-3 three-particle LCDAs with the three-particle decay constant $f_{3K}$.
     The shapes of the three-particle LCDAs are modelled using the parameters $\omega_{3K}$ and $\lambda_{3K}$.
     Twist-4 LCDAs are described in terms of the parameters $\delta^2_K$, $\kappa_{4K}$ and $\omega_{4K}$. All parameters in this category are renormalized at the scale $1\,\gev$.
     
     \item[\textbf{sum rule parameters and scales}] These include the Borel parameter $M^2$ and the duality threshold parameters
     for each of the form factors $\{f_+,f_0,f_T\}$.
     Since we use $q^2$ dependent duality thresholds, the latter involve the normalisation and slope of the threshold as functions of $q^2$,
     see the discussion in \autoref{sec:LCSR-duality}. The perturbative hard scattering kernels are evaluated at a renormalisation scale $\mu$,
     which is only varied a posteriori to assign a systematic uncertainty to the form factor calculations.
 \end{description}

 A summary of all input parameters and their prior probability density functions (PDFs) is presented in \autoref{tab:inputs}.
 We briefly discuss the differences between the inputs used in this work and the ones used previously in
 Refs.~\cite{Ball:2006wn,Duplancic:2008tk,Khodjamirian:2017fxg}: 
 \begin{enumerate}
     \item We update the value for the strange quark mass at our reference scale from $95 \pm 10~\mev$ to $93.4 \pm 8.6~\mev$.
     This change has a negligible effect on the numerical results.

     \item We use $a_{1K}$ and $a_{2K}$ from a recent lattice QCD analysis~\cite{LatticeParton:2022zqc}.

     \item We adapt the same Borel parameter window as in previous works. However, 
     contrary to those works, we do not apply a Gaussian approximation to the uncertainty arising from the Borel
     parameter. Instead, we use a uniform PDF as a prior, as done in Refs.~\cite{SentitemsuImsong:2014plu,Leljak:2021vte}.
 \end{enumerate}

\subsection{Duality thresholds}
\label{sec:LCSR-duality}

The duality thresholds $s_0^{f_i}$ represent splitting points which divide the dispersive integral for the corresponding form factors into two contributions: the $\bar{B}_s$ contribution, and the contribution due to excited $\bar{B}_s$ states and the continuum of $b\bar{s}$-flavoured states. A common procedure to constrain the duality thresholds is to use daughter sum rules. These are obtained by normalizing the derivatives of the form factors' correlation functions with respect to $-1/M^2$ to the correlation function
itself, yielding a $q^2$ estimator for the $B_s$ mass:
\begin{equation}
    \label{eq:lcsr:mass-estimator}
    \left[m_{B_s}^2(q^2;f_i)\right]_{\text{LCSR}} = \frac{\int_0^{s_0} \text{d}s\,s\,\rho^{f_i}(s,q^2)\,e^{-s/M^2}}{\int_0^{s_0} \text{d}s\,\rho^{f_i}(s,q^2)\,e^{-s/M^2}}\ .
\end{equation}
Here $f_i$ represents any of the form factors, and $\rho^{f_i}$ are the OPE results for its spectral density.\\

To determine the thresholds, we closely follow Ref.~\cite{Leljak:2021vte}:
First, we construct a Gaussian likelihood centered on the known $B_s$ mass.
This likelihood is a product of three uncorrelated likelihoods, one for each form factor.
Each likelihood involves the $q^2$-dependent predictions for the $B_s$ mass as obtained from the daughter sum rule.
We conservatively assign an uncertainty of $1\%$ for these theory predictions and impose the likelihoods'
constraints in five equally spaced $q^2$ points in the range $[-8~\gev^2,+8~\gev^2]$.\\
Second, we challenge the likelihood using two different models for the duality thresholds, as discussed below, and fit the model parameters
according to the priors in \autoref{tab:inputs}.
The posterior distributions of most parameters align well with their respective prior distributions.
The only exceptions are the duality threshold parameters and the Borel parameter, which are all distributed uniformly in the
prior.
Their posterior distributions exhibit a peaking behaviour, which indicates that we successfully inferred information
on both the duality thresholds and the Borel parameter from the likelihood.\\

\begin{figure}[t!]
  \begin{minipage}[c]{0.48\textwidth}
    \includegraphics[width=\textwidth]{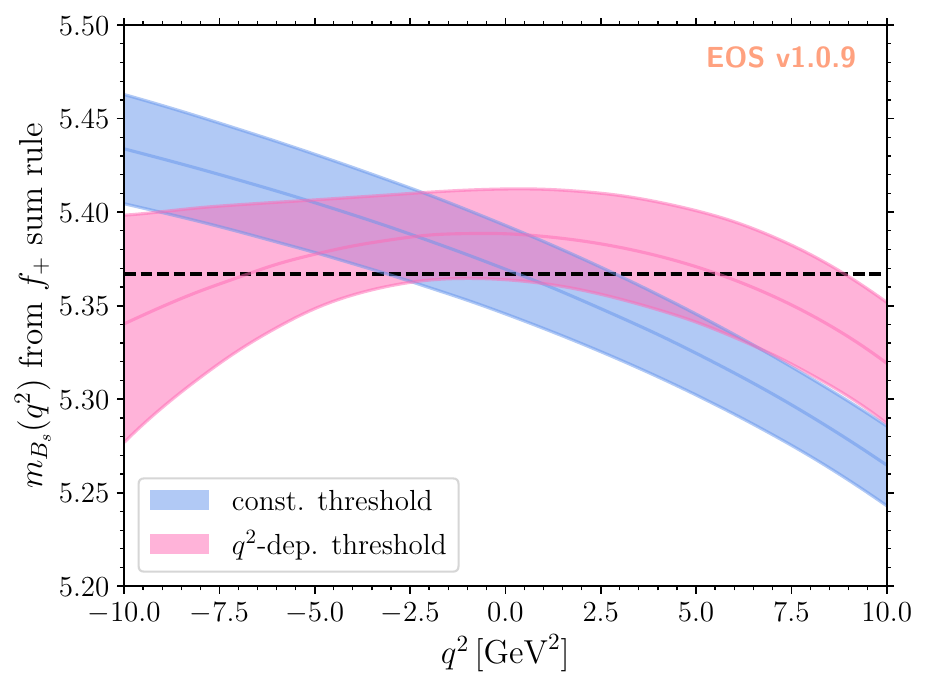}
  \end{minipage}\hfill
  \begin{minipage}[c]{0.48\textwidth}
    \includegraphics[width=\textwidth]{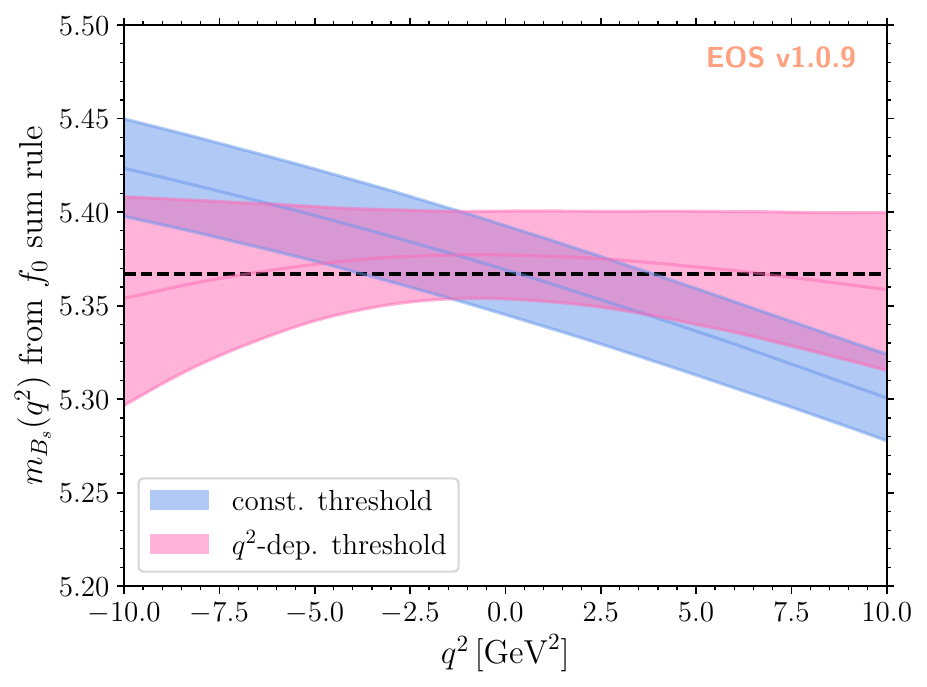}
  \end{minipage}
  
  \begin{minipage}[c]{0.48\textwidth}
    \includegraphics[width=\textwidth]{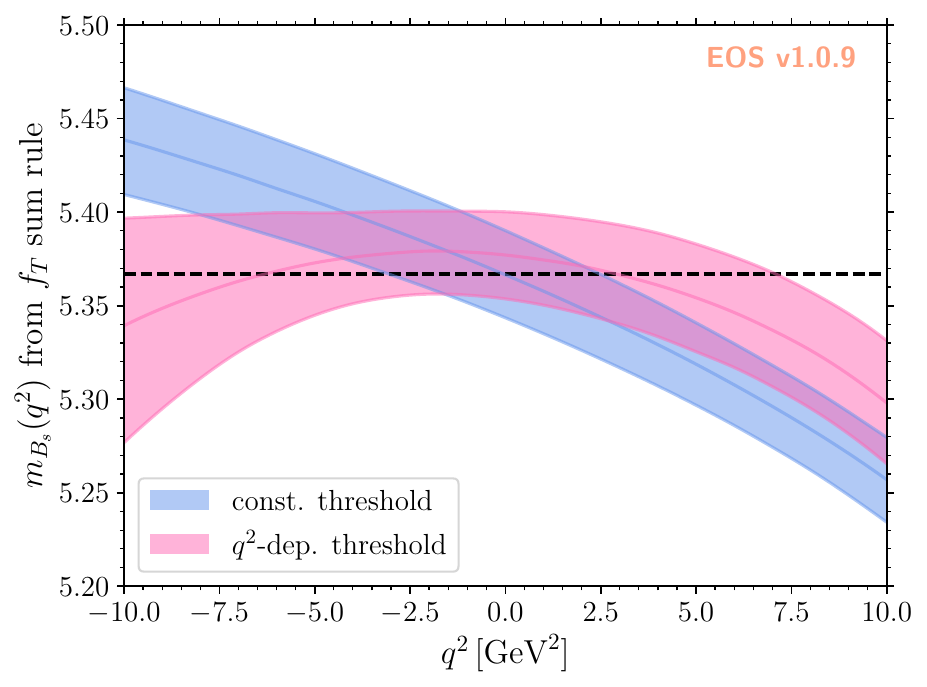}
  \end{minipage}\hfill
  \begin{minipage}[c]{0.48\textwidth}
    \caption{
       The $q^2$-dependence of the $B_s$-meson mass predictor $[m^2_{B_s}(q^2);f_i)]_{\textrm{LCSR}}$ for each of the three form factors $f_i = \{f_+, f_0, f_T \}$. The posterior-predictions for a $q^2$-invariant threshold~(blue) and a linearly dependent threshold~(pink) are shown. The shaded areas correspond to the respective 68\% probability envelopes. The dashed line corresponds to the known $B_s$ mass.
    } \label{fig:mBs-thresholds}
  \end{minipage}
\end{figure}

As in Ref.~\cite{Leljak:2021vte}, we use two models for the description of the duality thresholds:
$q^2$-independent thresholds; and $q^2$-dependent thresholds with a linear behaviour, $s_0^{f_i}(q^2) \equiv s_0^{f_i} + q^2\, s_0^{f_i}{}'$.
Similar to the situation for $\bar{B}\to \pi$ form factors~\cite{SentitemsuImsong:2014plu,Leljak:2021vte}, we find evidence for
a non-negligible $q^2$ dependence for the \bsk thresholds; the threshold values at the end of our
$q^2$ window vary by $\pm \sim 10\%$ compared to the values at $q^2 = 0$.
This observation is reflected in the overall fit quality: fitting the slope parameters reduces the $\chi^2$
of the theoretical likelihood from $8.2$ to $0.6$ at the expense of $3$ degrees of freedom.
The better fit is visible in \autoref{fig:mBs-thresholds}, where we plot the
$68\%$ probability envelopes of the estimators \autoref{eq:lcsr:mass-estimator} as functions of $q^2$
for either fit model. Despite the linear modelling of the $q^2$-dependence, the threshold-setting procedure is not able to
align the mass estimators for $f_+$ and $f_T$ with the known $B_s$ mass at $q^2 = 10\,\gev^2$.
We interpret this effect as a breakdown of the LCOPE for the underlying correlators.
Hence, we abstain from predicting any of the form factors at $q^2 > 5\,\gev^2$.

\subsection{Numerical results for LCSR form factors}
\label{sec:LCSR-formfactors}

The form factors are obtained by producing posterior-predictive samples at 4 equally-distanced $q^2$
points in the interval $-10~\gev^2 \leq q^2 \leq + 5~\gev^2$.
The samples for $f_0(0)$ are not included. They coincide, by construction, with those for $f_+(0)$ and would yield a singular covariance matrix if used. The form factors are evaluated using
the threshold model with linear $q^2$ dependence.
We obtain form factor samples that follow to good approximation a multivariate gaussian distribution.
This leads us to infer the form factors' values and covariance matrix by means of an
unbinned multivariate gaussian fit to the samples.

We account for systematic uncertainties by varying the renormalisation scale $\mu$
by $25\%$ of its baseline value, corresponding to the range
$\mu \in [2.40~\gev, 3.75~\gev]$. We then evaluate the form factors for the central values of
the remaining input parameters. We find that lowering the renormalisation scale incurs the numerically
largest shift in the form factors, corresponding to a maximal decrease of the central
values by $5.0\%$. We account for this systematic uncertainty through a diagonal covariance matrix,
with entries corresponding to the square of the maximal shifts of the central values,
\begin{equation}
    \Sigma_{ii} \big|_{\mu} = \max(|f_i(\mu_\text{low}) - f_i|, |f_i(\mu_\text{hi}) - f_i|)^2\ .
\end{equation}
In addition, we add a systematic uncertainty for the threshold model. To do so, 
we produce form factor samples using the threshold model without $q^2$ dependence.
We obtain their mean values across all $q^2$ points. We assign a systematic uncertainty based on the square of
their difference to the nominal form factors results. The corresponding covariance matrix is strictly populated on
the diagonal,
\begin{equation}
    \Sigma_{ii} \big|_\text{thr.} = |f_i^\text{const} - f_i^\text{$q^2$-dep.}|^2\ .
\end{equation}
We find that the biggest source of systematic uncertainty comes from changing the threshold model.
Switching from a $q^2$-dependent to
a $q^2$-independent threshold model we find shifts to the central form factor values ranging from
$0.7\%$ to $12.2\%$.
The total covariance matrix is then obtained as the sum of the parametric covariance matrix with both systematic
covariance matrices:
\begin{equation}
    \Sigma \big|_\text{total} = \Sigma \big|_\text{param} + \Sigma \big|_{\mu} + \Sigma \big|_\text{thr.}
\end{equation}

We approximate the joint posterior predictive distribution of all form factors at the different $q^2$ points
as a multivariate Gaussian distribution. The mean values and standard deviations of the LCSR predictions are given in \autoref{tab:results:LCSR}. 
The central values and total covariance matrix are provided as a machine-readable file as part
of the ancillary material~\cite{EOS-DATA-2023-03} and within the \EOS software as of v1.0.9 as a constraint labelled
\begin{center}
    \texttt{B\_s->K::form-factors[f\_+,f\_0,f\_T]@BvDV:2023A}
\end{center}
Our results are compatible with the previous LCSR results for $f_+(q^2 = 0)$ in \autoref{eq:intro:fp-DM2008} and \autoref{eq:intro:fp-KR2017}
at the $1.3\,\sigma$ and $0.8\,\sigma$ levels, respectively.
A meaningful comparison with the full set of results in Refs.~\cite{Duplancic:2008tk,Khodjamirian:2017fxg}
is not possible, due to their lack of correlation information across different form factors.
We find the relatively largest systematic uncertainty in our results at $q^2 = -10\,\gev^2$.
At this point, total uncertainties for the form factors vary between $17\%$ and $19\%$.
At larger values of $q^2$, the relative uncertainties are significantly smaller, ranging all consistently
between $6\%$ and $9\%$.
We find that our procedure to account for systematic uncertainties significantly decorrelates
our results. The average degree of correlation, assuming dominant correlation to nearest neighbours,
decreases from $\sim 76\%$ to $\sim 40\%$.

\begin{table}[t!]
    \centering
    \begin{tabular}{c c c c c}
        \toprule
        $q^2$       & $-10\;\gev^2$         & $-5\;\gev^2$         & $0\;\gev^2$          & $+5\;\gev^2$ \\
        \midrule
        $f_+(q^2)$  & $0.208 \pm0.035$      & $0.278\pm0.022$      & $0.364\pm0.026$      & $0.482\pm0.042$\\
        
        $f_0(q^2)$  & $0.261 \pm0.047 $     & $0.312\pm0.025$      & \textemdash          & $0.425\pm0.040$\\
        
        $f_T(q^2)$  & $0.232 \pm 0.044$     & $0.305\pm0.027$      & $0.394\pm0.023$      & $0.516\pm0.035$ \\
        \bottomrule
    \end{tabular}
    \caption{%
        Our results for the \bsk form factors from the LCSR analysis.
        The central values arise from a weighted average of the posterior-predictive samples,
        and the uncertainties arise from a combination of the parametric and systematic uncertainties; see text.
        The result for $f_0(0)$ has been omitted due to the identity $f_+(0)=f_0(0)$.
    }
    \label{tab:results:LCSR}
\end{table}

\section{Form factors in the full $q^2$ range}
\label{sec:ff}

\subsection{Parametrisation}
\label{sec:ff:param}
To access the full semileptonic range in $q^2$, we have to apply a parametrisation for the
interpolation or extrapolation of the available form factor data. Common parametrisations that are used respect the analyticity and unitarity properties of the form factors, like the BGL parametrisation used in $\bar{B}\to D^{(*)}$ form factors~\cite{Boyd:1994tt} (see~\cite{Caprini:2019osi} for a textbook discussion). Most importantly, the form factor is expanded in powers of $z(q^2)$, which conformally
maps the form factor's first Riemann sheet onto the open unit disk in the complex $z$ plane: 
\begin{equation}
    q^2 \mapsto z(q^2; t_\Gamma^f, t_0^f)
        =   \frac{%
                \sqrt{t_\Gamma^f - q^2}
                -
                \sqrt{t_\Gamma^f - t_0^f}
            }{%
                \sqrt{t_\Gamma^f - q^2}
                +
                \sqrt{t_\Gamma^f - t_0^f}
            }
\end{equation}
Here, $t_0^f$ is a free parameter that is used to fix the zero crossing
of $z(q^2 = t_0^f) = 0$, and $t_\Gamma^f$ represents the first
branch point of the form factor $f$. Throughout this work we use $t_0^f = 14.7\,\gev^2$.

For the \bsk form factors, we then parametrize the full form factor as
\begin{equation}\label{eq:modBGL}
    f(q^2) = \frac{1}{\sqrt{\chi_f} \phi_f(q^2) B_f(q^2)} \sum_k^K a_k^f p_k(z(q^2))\ ,
\end{equation}
where $p_k$ are a suitable choice of polynomials discussed below.
The quantities $\chi_f$ and $\phi_f$ are known; they arise from the computation of the
unitarity bound within an operator product expansion of a suitable correlation function. 
 The quantity $B_f(q^2)$ accounts for a finite number of isolated poles beyond the semileptonic phase space but below the first branchpoint of the function. 

In the traditional BGL-like setup, the first branch point $t_\Gamma^f$ of the form factor $f$
coincides with the pair production threshold of the process $B\to P$, $t_+ = (m_{B} + m_{P})^2$,
where $P$ respresent a pseudoscalar state. In that case, the polynomials $p_k$ reduce to
$z^k / \sqrt{2\pi}$.
However, the $\bsk$ form factors develop their first branch point at $t_\Gamma = (m_B + m_\pi)^2$,
since from this point forward on-shell $\bar{B}\pi$ states can rescatter into $\bar{B}_s\bar{K}$ states.
This branch point does not coincide with the pair production threshold
$t_+ = (m_{B_s} + m_{K})^2$, thereby breaking a central assumption of the BGL approach.

The parameterisation in \autoref{eq:modBGL} accounts for this mismatch by considering the pair-production cross section only in the interval $[(m_{B_s} + m_{K})^2, \infty)$, rather than
starting at the first branch point. The unitarity bound ensures that
\begin{align}
    \label{eq:ff:param:boundary-integral}
    \oint_{z \in \mathcal{Z}} \frac{dz}{z}
    \bigg|\phi_f B_f f\bigg|^2_{q^2 = q^2(z)} \leq 1\ ,
\end{align}
where the integration domain now only covers an arc of the unit circle in the complex $z$ plane:
\begin{equation}
    \label{eq:ff:param:domain-BGL}
    \mathcal{Z} = \lbrace z \, | \, |z| = 1 \land |\arg z| \leq |\arg z(t_+)|\rbrace \ .
\end{equation}
BGL-like parametrisations that respect this type of unitarity bound have recently been
developed in applications to $\Lambda_b \to \Lambda^{(*)}$ form factors~\cite{Blake:2022vfl,Amhis:2022vcd},
$\bsk$ form factors~\cite{Flynn:2023qmi}, and $\bar{B}_{(s)}\to \lbrace \bar{K}^ {(*)},\phi\rbrace$ form factors~\cite{Gubernari:2023puw}; and non-local form factors
in $\bar{B}\to \bar{K}\gamma^*$ transitions~\cite{Gubernari:2020eft,Gubernari:2022hxn}.
Here, we use the approach first discussed in Ref.~\cite{Gubernari:2020eft}. In this case, the polynomials $p_k$ are orthonormal with respect to the measure $dz/z$ on the
integration domain \autoref{eq:ff:param:domain-BGL}. They can be efficiently computed using
the Szeg\H{o} recurrence relation; we refer to appendix B of Ref.~\cite{Gubernari:2022hxn} for details.
An alternative approach exists, which diagonalizes the bound a-posteriori~\cite{Flynn:2023qmi}.
We emphasize that both approaches yield identical results.
A numerical implementation of \autoref{eq:modBGL} including the polynomials, the outer functions, 
and the Blaschke factors is available in the \EOS software and documented
in Ref.~\cite{Gubernari:2023puw}. We use this implementation and truncate the series at order $K=4$,
which is compatible with the observation that such a high truncation order
is required to stabilize the extrapolation to $q^2=0$~\cite{Flynn:2023qmi,Flynn:2023nhi}.\\

The perturbative component of the unitarity bound is encoded in the
numerical values for the quantity $\chi_f$. It is obtained from a subtracted
dispersion relation for a suitable vacuum matrix element of a two-point
correlation function that involves two insertions of $b\to u$ currents. We apply isospin symmetry to relate the values for $\chi_f$ in $b \to d$
processes provided in Ref.~\cite{Bharucha:2010im} to obtain the values for $\chi_f$ in $b\to u$ currents
required here. 
For convenience, we provide the values used for the three form factors
discussed here:
\begin{equation}
\begin{aligned}
    \chi^{f_+}
        & = 6.58\times 10^{-4} \;\gev^{-2}\ , &
    \chi^{f_0}
        & = 1.50 \times 10^{-2}\ , &
    \chi^{f_T}
        & = 4.39\times 10^{-4} \;\gev^{-2}\ .
\end{aligned}
\end{equation}
We do not include any isospin breaking corrections in our approximation, which are of order $(m_u - m_d) / m_b$ and $\alpha_e/\pi$.
We further sharpen the bounds by accounting for the polarisation
of the intermediate $\bar{B}_s\bar{K}$ pair following Ref.~\cite{Gubernari:2023puw}.
The effect of this approach
is a rescaling of the perturbative values $\chi^f$ for the form factors
$f_+$ and $f_T$.
In this regard, our work goes beyond what has been done in Ref.~\cite{Flynn:2023qmi}.

\subsection{Analyses of the available form factor data}
\label{sec:ff:results}

\begin{figure}[t!]
  \begin{minipage}[c]{0.48\textwidth}
    \includegraphics[width=\textwidth]{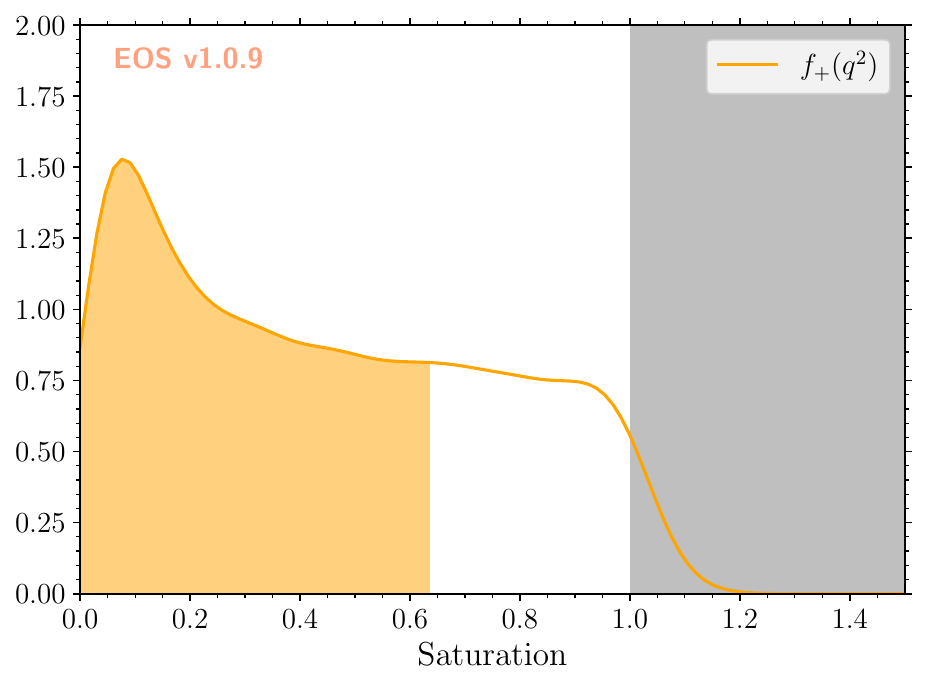}
  \end{minipage}\hfill
  \begin{minipage}[c]{0.48\textwidth}
    \includegraphics[width=\textwidth]{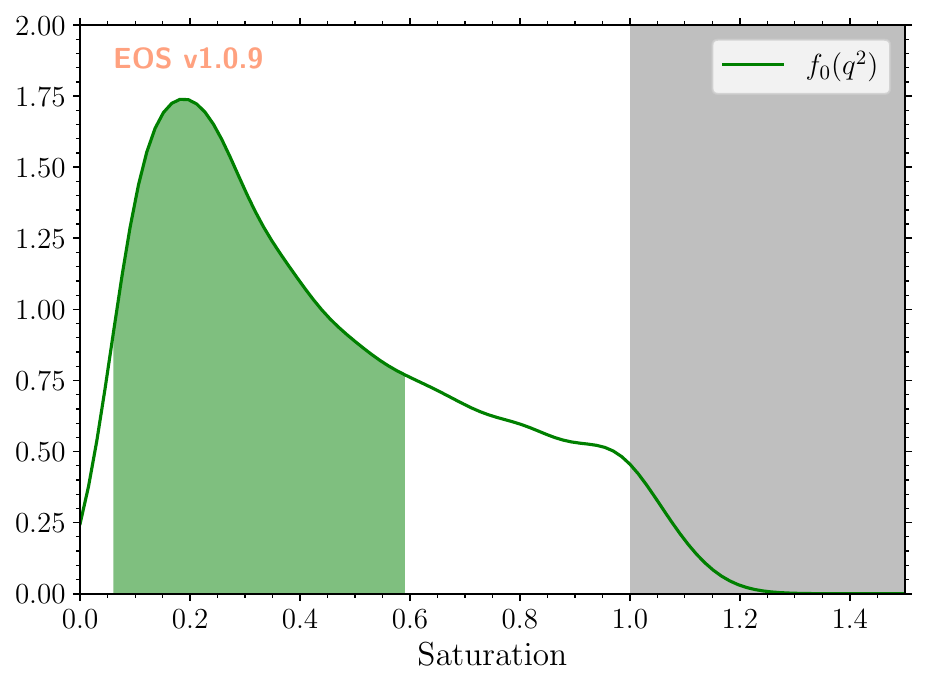}
  \end{minipage}
  
  \begin{minipage}[c]{0.48\textwidth}
    \includegraphics[width=\textwidth]{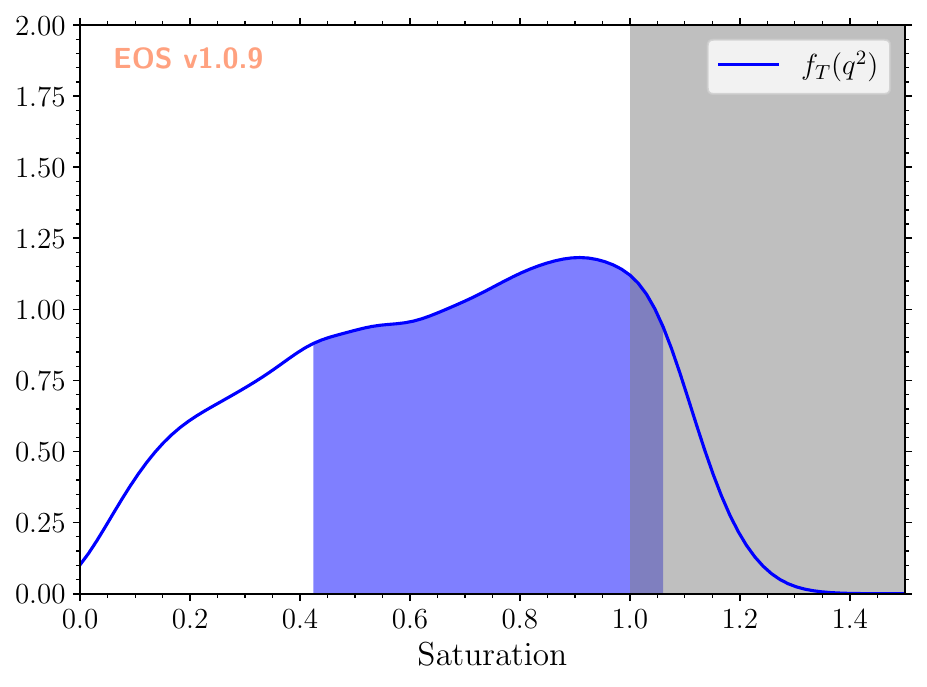}
  \end{minipage}\hfill
  \begin{minipage}[c]{0.48\textwidth}
    \caption{%
        Saturation $\text{sat}_i$ of the unitarity bound for each form factor $f_i$ for a truncation order $K=4$.
        See \autoref{eq:ff:saturations} for their definition. We show a continuous histogram of the
        posterior-predictive PDF for each saturation, with their respective $68\%$ intervals
        shown as the coloured areas.
        The grey shaded areas represent saturations that exceed the
        value allowed by the unitarity bound.
    }
    \label{fig:ff:saturation}
    \end{minipage}
\end{figure}

Throughout, we truncate the parametrisation of the form factors at order $K = 4$,
which corresponds in general to $5$ parameters per form factor.\footnote{%
    The priors for analyses that do not include data on the tensor form factor
    are restricted to the parameters for $f = f_+, f_0$ only.
}
As prior we use a product of independent uniform PDFs for each of the free form factor parameters $a_k^f$,
$0 \leq k \leq K$, with support $-1 \leq a_k^f \leq +1$.
Note that the parameter $a_0^{f_0}$ is not a free parameter; instead,
it is fixed so as to fulfill the identity $f_+(0) = f_0(0)$.\\

We further define the named likelihoods:
\begin{description}
    \item[\textbf{LQCD}] This likelihood contains the available lattice QCD results on \bsk form factors
    $f_+$ and $f_0$ by the HPQCD~\cite{Bouchard:2014ypa} and RBC/UKQCD~\cite{Flynn:2023nhi} collaborations.
    We directly use the $2+3$ synthetic data points provided by RBC/UKQCD for $f_+$ and $f_0$, respectively,
    including their covariance matrix.
    We further produce $3+3$ synthetic data points for the HPQCD results, using the same $q^2$ values as RBC/UKQCD%
\footnote{%
    We find that our fit results in this section, the phenomenological results
    in \autoref{sec:pheno}, and our conclusions are stable with respect to systematic
    shifts of the synthetic HPQCD data points by $-1\,\gev^2$ and $-2\,\gev^2$.
}.
    We do not use the results by the FNAL/MILC collaboration~\cite{FermilabLattice:2019ikx} due to a suspected issue
    with the chiral extrapolation; see the corresponding discussion in the conclusion of Ref.~\cite[p. 21]{Flynn:2023nhi}.
    This likelihood hence corresponds to a total of $11$ observations.

    \item[\textbf{LCSR}] This likelihood contains our synthetic data points obtained from the light-cone sum rule analysis
    that we carry out in \autoref{sec:lcsr}. We use a total of $4$ points in $q^2$ for both the $f_+$ and $f_T$ form factors
    and $3$ points in $q^2$ for the $f_0$ form factor.
    This likelihood hence contributes an additional $11$ observations.
\end{description}

We define a total of three posterior PDFs labelled \textbf{LCSR}, \textbf{LQCD}, and \textbf{LCSR+LQCD}.
They use the common prior and one of the likelihoods or the product of both likelihoods, corresponding to their label.
The posteriors labelled LCSR and LQCD are underconstrained. This is manifest for the LCSR posterior,
since the number of parameters ($14$) exceeds the number of observations ($11$).
In the case of the LQCD posterior, the choice to use the same $q^2$ values to generate synthetic data points
for the HPQCD results leads to only \emph{$6$ effective} observations for $9$ parameters.
Both cases can only be meaningfully analysed due to the application of the unitarity bound,
which is built into the prior PDF in its weakest form, restricting the domain of the posterior PDF
to a hypercube.
In addition, we apply the unitarity bound in a slightly stronger form as follows.
For each point in the parameter space, we compute the three saturations
\begin{equation}
\label{eq:ff:saturations}
\begin{aligned}
    \text{sat}_+
        & \equiv \sum_k |a_k^{f_+}|^2 \ , &
    \text{sat}_0
        & \equiv \sum_k |a_k^{f_0}|^2 \ , &
    \text{sat}_T
        & \equiv \sum_k |a_k^{f_T}|^2 \ .
\end{aligned}
\end{equation}
The unitarity bounds limit each saturation to $1$.
For points that exceed a saturation of $1$, we penalize each posterior PDF $P$ with a half-gaussian term
\begin{equation}
    \label{eq:ff:penalty-function}
    \log P \supset \sum_{i=+,0,T} \begin{cases}
        -\frac{1}{2} \left(\frac{\text{sat}_i - 1}{\sigma_i}\right)^2 & \text{if } \text{sat}_i \geq 1\\
        0 & \text{otherwise}
    \end{cases}\ ,
\end{equation}
as suggested in Ref.~\cite{Bordone:2019vic} in the context of $b\to c$ form factor bounds.
Here $\sigma_i$ represents the relative uncertainty on the quantities $\chi^{f_i}$; including
this uncertainty in the description somewhat loosens the bounds.
We use $\sigma_i = 10\%$ for all $i \in \lbrace+, 0, T\rbrace$, which corresponds to the relative uncertainties for the quantities $\chi^f$ obtained in Ref.~\cite{Bharucha:2010im}.

Our statistical analysis is carried out using the \EOS software~\cite{EOSAuthors:2021xpv}
in version v1.0.9~\cite{EOS:v1.0.9}. As part of our analysis, we draw importance samples
from the three posterior PDFs. For this task, we rely on the \dynesty software~\cite{Speagle:2020,dynesty:v2.0.3}
to produce these importance samples using dynamical nested sampling~\cite{Higson:2018}.\\

We maximize the three posterior densities with respect to the form factor parameters. In the case of the underconstrained posteriors LCSR and LQCD, this leads to multiple solutions in the parameter space that share the same minimal $\chi^2$ value. In the case of the posterior LCSR+LQCD, the optimisation yields an isolated best-fit point.
We provide an overview of the $\chi^2$ values at the best-fit points
in \autoref{tab:ff:results}, with the following caveats: 
\begin{itemize}
    \item The fit to the LCSR data has negative degrees of freedom,
    which makes a goodness-of-fit check based on the $\chi^2$ test statistic impossible.

    \item The fit to the LQCD data has effectively negative degrees of freedom, since the
    two individual likelihoods do not provide complementary information.
    Despite this, the $\chi^2$ value in the best-fit point is expected to be non-zero,
    since it effectively represents the goodness of fit of the weighted average of the
    two LQCD likelihoods for only $5$ degrees of freedom.

    \item The penalty term \autoref{eq:ff:penalty-function} does not enter the likelihood
    and can be understood as a prior density. Moreover, we have no appropriate test statistics
    for the penalty term. As a consequence, we do not account for the penalty term in the
    goodness-of-fit discussion.
\end{itemize}
 We find that our nominal LCSR+LQCD posterior provides
an acceptable fit, since its $p$-value of $\sim 6\%$ exceeds our a priori threshold of $3\%$.
We also find that the two LQCD likelihoods are in good agreement with each other:
their weighted average yields $\chi^2 / \text{d.o.f} = 5.7 / 5$, which corresponds to a
$p$-value of $\sim 34\%$.

\begin{table}[t!]
    \centering
    \begin{adjustbox}{center}
    \begin{tabular}{c D{.}{.}{-1} D{.}{\ }{-2} D{.}{.}{-1} c c c c c}
        \toprule
        ~
            & \multicolumn{3}{c}{goodness of fit}
            & \multicolumn{3}{c}{BFP~saturation}
            & \multicolumn{2}{c}{extrapolation}\\
        posterior
            & \multicolumn{1}{c}{$\chi^2$}
            & \multicolumn{1}{c}{d.o.f.}
            & \multicolumn{1}{c}{$p$-value}
            & $\text{sat}_+$
            & $\text{sat}_0$
            & $\text{sat}_T$
            & $f_+(q^2=0)$
            & $f_T(q^2=0)$\\
        \midrule
        LCSR
            & 0.0
            & -3
            & \multicolumn{1}{c}{\textemdash}
            & 0.93
            & 1.00
            & 1.00
            & $0.36 \pm 0.02$
            & $0.39 \pm 0.02$\\
        \midrule
        LQCD
            & 5.7
            & -3
            & \multicolumn{1}{c}{\textemdash}
            & 0.45
            & 0.52
            & \textemdash
            & $0.25 \pm 0.08$
            & \textemdash\\
        \midrule
        LCSR+LQCD
            & 15.0
            & 8
            & 6.0\%
            & 1.01
            & 0.34
            & 1.00
            & $0.31 \pm 0.02$
            & $0.36 \pm 0.02$\\
        \bottomrule
    \end{tabular}
    \end{adjustbox}
    \caption{Comparison of the three fits.
    We provide goodness-of-fit diagnostics like the $\chi^2$ in the best-fit point (BFP),
    the degrees of freedom (d.o.f.) and the $p$-value (where applicable, see the text for
    a discussion).
    We further provide the saturation of the unitarity bounds in the respective BFPs.
    Finally, we provide the form factors $f_+$ and $f_T$ at $q^2=0$.
    }
    \label{tab:ff:results}
\end{table}

We observe that the unitarity bounds affect the fits substantially.
As shown in \autoref{tab:ff:results}, the LCSR best-fit points violate the unitarity bounds with saturations close to or exceeding unity
in all three form factors. The LQCD best-fit points respect the unitarity bounds, with saturations of $\sim 50\%$ for the $f_+$
and the $f_0$ bounds. The best-fit point of our nominal LCSR+LQCD fit shows a violation of the unitarity bounds for $f_+$ and $f_T$.

Moreover, we find that the posterior samples readily saturate the unitarity bounds.
This is illustrated in \autoref{fig:ff:saturation}, where we show the posterior-predictive PDF
for the saturations.
This broad distribution for the saturation of the bounds is expected, given the large number of fit parameters.
As a consequence, we find that the distribution of the fit parameters does not resemble
a multivariate gaussian distribution. While individual marginal posterior densities
look gaussian-like, the joint distribution is highly distorted due to the effect of the
unitarity bounds. Hence, unlike in our fit to the LCSR results,
we do not carry out an unbinned fit to the posterior samples.

We find that the unitarity bounds in the BFP in~\autoref{tab:ff:results} for the LCSR and LCSR+LQCD posteriors are \emph{effectively saturated}\footnote{%
    We remind here that we do not include other transitions such as $\bar{B}\to \lbrace \pi,\rho,\omega\rbrace$
    when computing the saturation of the unitarity bounds.
}, thereby violating unitarity.
However, the full distribution of the posterior samples covers a substantial range of smaller saturation values as seen in \autoref{fig:ff:saturation}.
Therefore, we do not consider the over saturation of the BFP a sufficient criterion to discard these fit results.

\begin{figure}[t!]
  \begin{minipage}[c]{0.48\textwidth}
    \includegraphics[width=\textwidth]{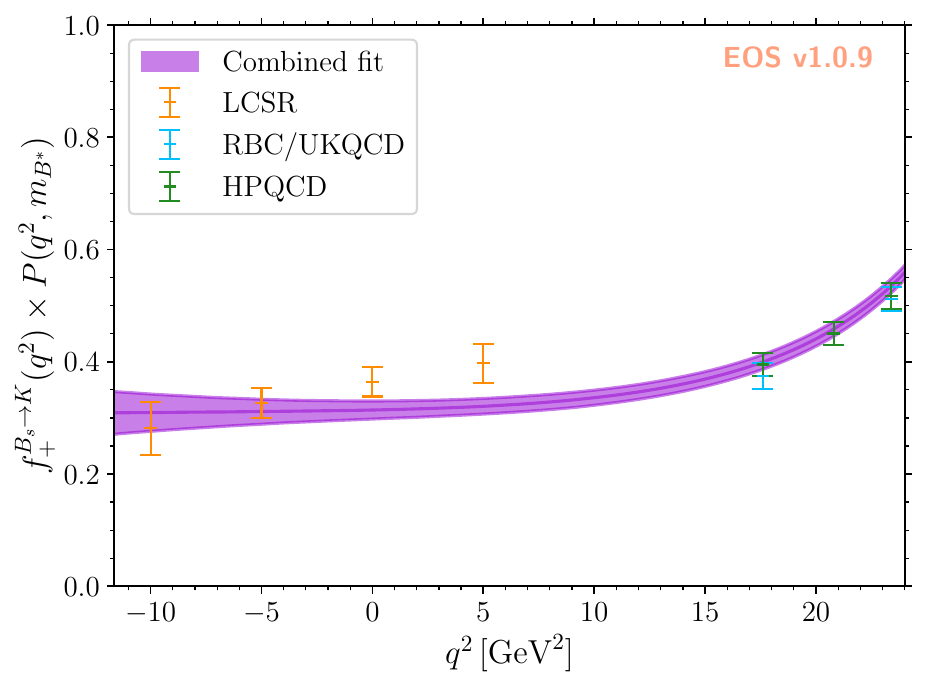}
  \end{minipage}\hfill
  \begin{minipage}[c]{0.48\textwidth}
    \includegraphics[width=\textwidth]{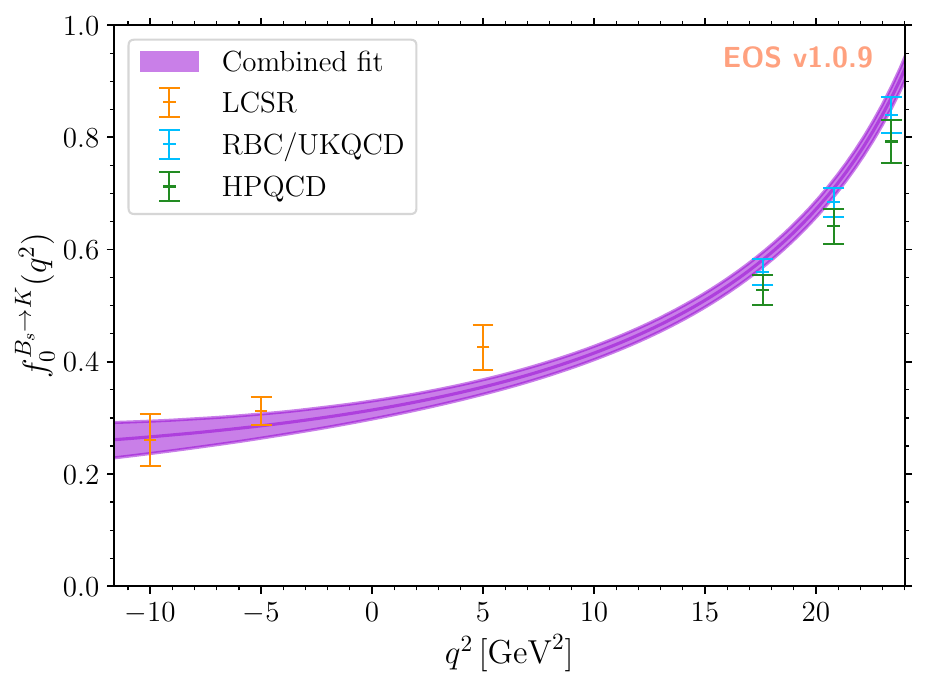}
  \end{minipage}
  
  \begin{minipage}[c]{0.48\textwidth}
    \includegraphics[width=\textwidth]{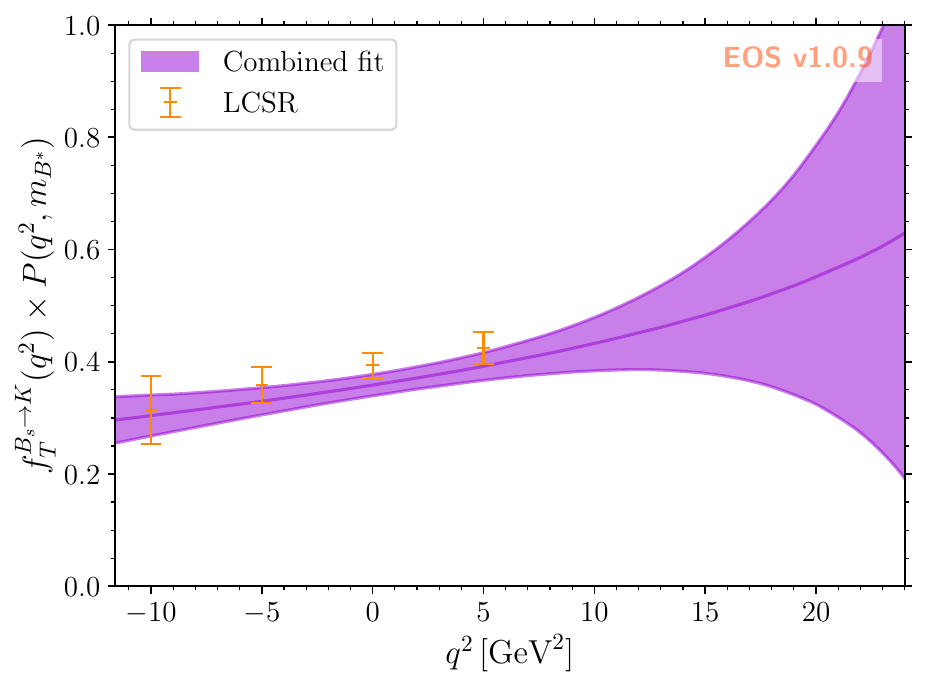}
  \end{minipage}\hfill
  \begin{minipage}[c]{0.48\textwidth}
    \caption{
       Results of the simultaneous fit to all form factors for a truncation order $K=4$, combining LCSR and LQCD constraints.
       The shaded bands correspond to the $68\%$ probability envelopes of our posterior predictions.
       The coloured crosses indicate the various experimental and statistical constraints and their uncertainties
       as described in the text.
    } \label{fig:ff:results}
  \end{minipage}
\end{figure}

Due to the non-gaussianity of the samples, we use the posterior samples to produce posterior-predictive
samples for the three form factors at various $q^2$ points, including at $q^2 = 0$. The latter results are given in \autoref{tab:ff:results}.
For the LCSR+LQCD posterior, we show the median values and central $68\%$ probability envelopes for the form 
factors as a function of $q^2$ in \autoref{fig:ff:results}. Data points for LCSR and LQCD likelihoods are also shown.
Further plots illustrating the differences amongst the three posteriors
are available in the supplementary material~\cite{EOS-DATA-2023-03}.
We emphasize that the accurate estimation of the form factors uncertainties requires the use of the posterior samples,
which we also make available as part of the supplementary material.

Based on the above considerations, we use the results of the LCSR+LQCD fit as our nominal fit results.

\begin{figure}[t!]
    \centering
    \includegraphics[width=0.85\textwidth]{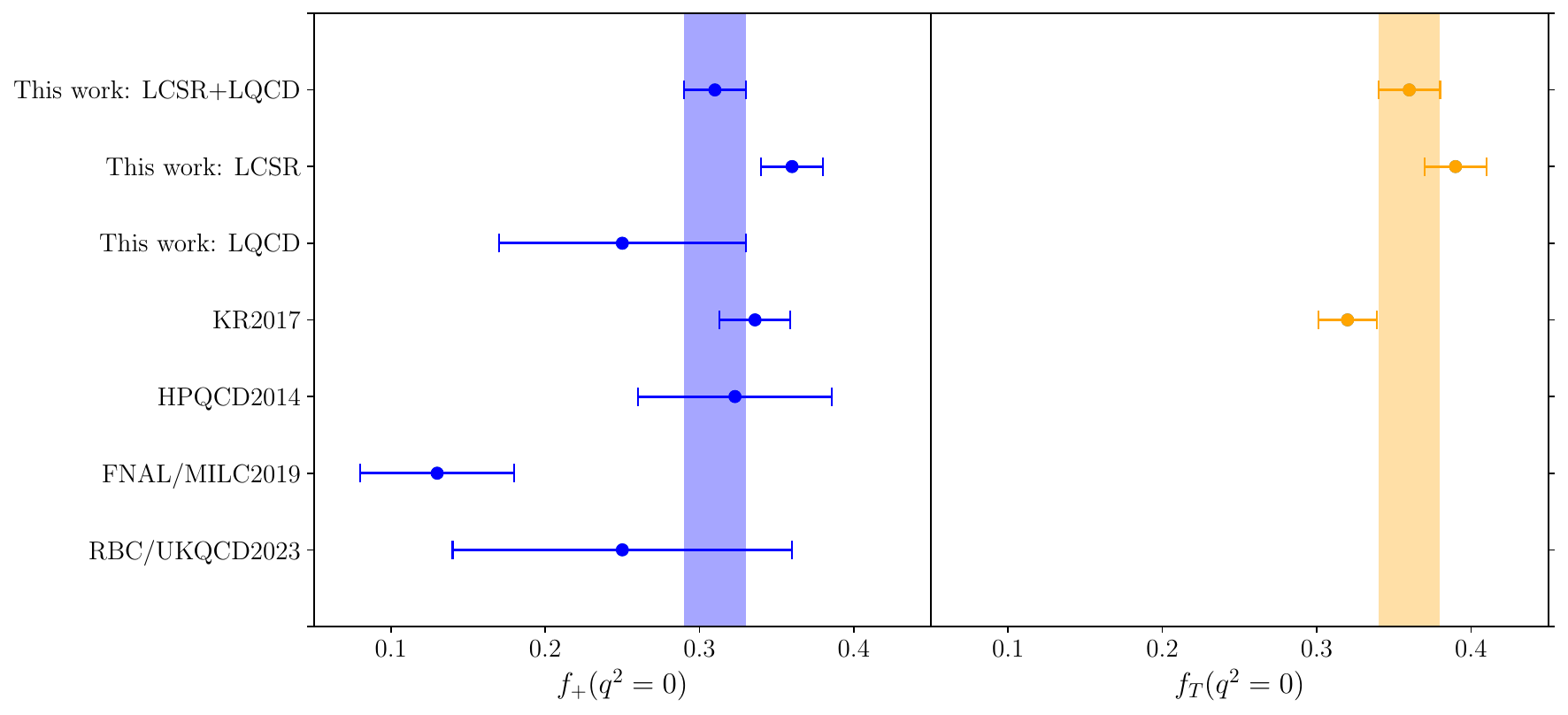}
    \caption{
       Comparison of our determinations of $f_+(q^2=0)$ and $f_T(q^2=0)$ with the inputs and the literature.
       The bands present the $68\%$ probability interval of our nominal results based on the LCSR+LQCD posterior.
    } \label{fig:ff:compare}
\end{figure}

Finally, we compare our results at $q^2=0$ GeV$^2$ with the different LCSR and LQCD results discussed in the introduction as shown in \autoref{fig:ff:compare}. For completeness, we also include the FNAL/MILC2019 determination, which is not included in our fit, as discussed above.

\section{Phenomenology}
\label{sec:pheno}
\subsection{Differential decay rate}

\begin{figure}[t!]
    \centering
\begin{minipage}{.5\textwidth}
  \centering
    \includegraphics[width=0.95\textwidth]{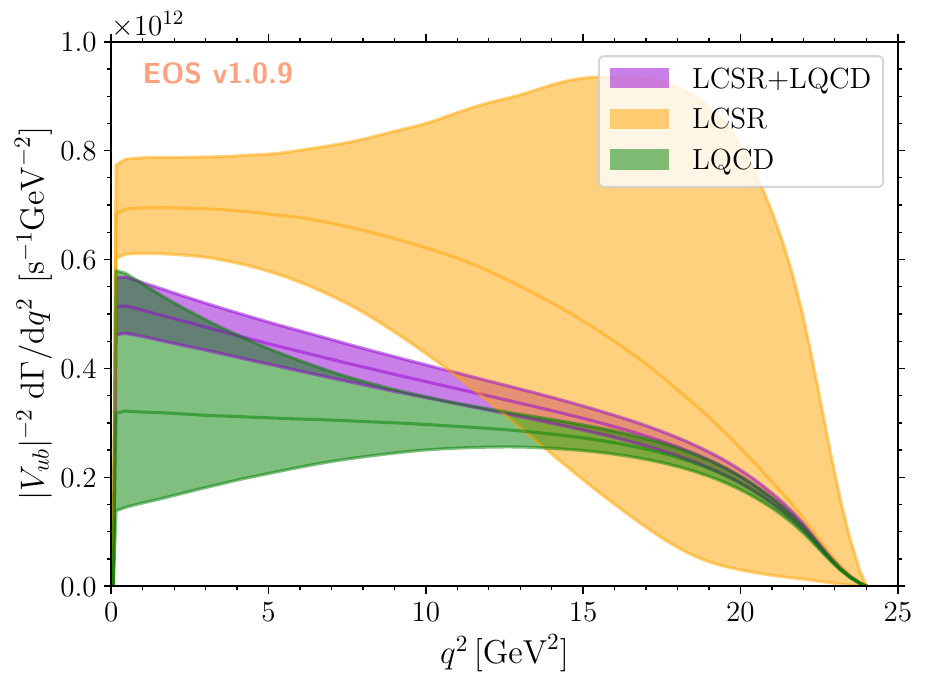}
\end{minipage}%
\begin{minipage}{.5\textwidth}
  \centering
    \includegraphics[width=0.95\textwidth]{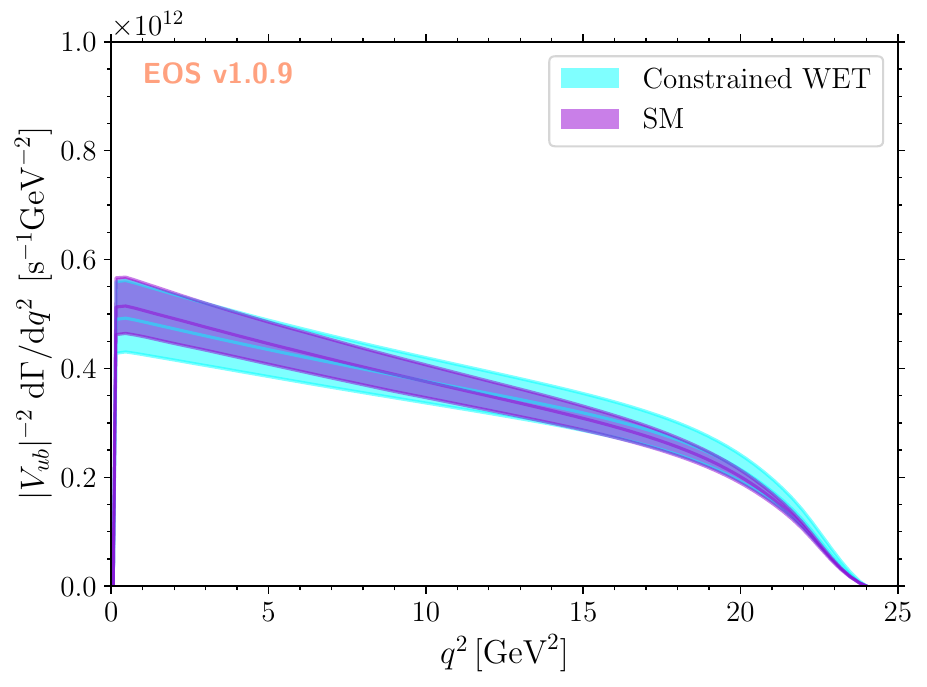}
\end{minipage}
    \caption{%
        Differential decay rate of $\bar{B}_s^0\rightarrow K^+ \mu^- \bar{\nu}$
        as a function of the momentum transfer $q^2$
        for the three different analyses within the SM~(left) and in comparison to our nominal
        fit for the allowed BSM reach~(right).
    }
    \label{fig:pheno:plot}
\end{figure}

We can now determine the differential decay rate of $\bar{B}_s^0\rightarrow K^+ \mu^- \bar{\nu}$ in units of $|V_{ub}|^2$.
In the next section, we use specific bins of this distribution to determine the ratio $|V_{ub}/V_{cb}|$ from experimental data.
However, we stress that the shape of the distribution gives additional information which should be confronted with experimental data.
In \autoref{fig:pheno:plot} (left), we show the obtained posterior-predictions for the differential decay rate of $\bar{B}_s^0\rightarrow K^+ \mu^- \bar{\nu}$ for our nominal fit (purple) and separately for the LCSR (yellow) and LQCD (green) posterior predictions.
We observe that the LCSR (LQCD) determination is---as expected---most precise at low (high) $q^2$.
In both cases, the unitarity bounds limit the uncertainty. We stress that these two determinations are compatible; the $p$-value of the combined fit is $6\%$ (see \autoref{sec:ff:results}), and at low $q^2$ the two bands are compatible at less than two standard deviations. Finally, we note that our combined fit has smaller uncertainties in every $q^2$ point than the smallest uncertainty in every individual fit, another indication that the two sets of information are mutually compatible.

As pointed out in Ref.~\cite{Jung:2018lfu}, scalar contributions beyond the Standard Model (BSM) have the potential
to significantly distort the shape of the $q^2$ distribution in $P \to P\ell\bar\nu$ decays.
A recent study of $\bar{B}\to \lbrace\pi,\rho,\omega\rbrace \ell^-\bar{\nu}$ decays \cite{Leljak:2023gna}, which are mediated by the $b\to u\ell^-\bar\nu$ transition,
constrains the available parameter space of the beyond the Standard Model (BSM) Wilson coefficients,
specifically, allowing for new scalar, tensor and left and right-handed vector interactions.
In \autoref{fig:pheno:plot} (right), we illustrate the BSM reach of the differential $\bar{B}_s^0\to K^+\ell^-\bar\nu$ distribution, combining our new form factor results with the BSM parameter samples provided in Ref.~\cite{Leljak:2023gna,EOS-DATA-2023-01v2}. We observe that $\bsk$ has similar sensitivity to BSM parameters as the $\bar{B}\to (\pi, \rho)$ transitions, resulting in little room left for BSM contributions.
The distribution, however, is shifted slightly and allows for more $\bar{B}_s^0\to K^+\ell^-\bar\nu$ events at high $q^2$ than in the SM.
However, at the current level of precision we observe no significant distortion of the distribution.

\subsection{\boldmath Determination of $|V_{ub}/V_{cb}|$}
The LHCb collaboration recently observed the $B_s^0\to K^- \mu^+\nu_\mu$ decay for the first time \cite{LHCb:2020ist}.
Its integrated branching ratio is obtained as follows:
\begin{equation}
    R_{\rm BF} \equiv \frac{\mathcal{B}(B_s^0\to K^- \mu^+\nu_\mu)}{\mathcal{B}(B_s^0\to D_s^- \mu^+\nu_\mu)} = \frac{N_K}{N_{D_s}}\frac{\epsilon_{D_s}}{\epsilon_K} \times \mathcal{B}(D_s^-\to K^+K^-\pi^-) \ ,
\end{equation}
where $\epsilon_X$ are the efficiencies. The yields for $B_s^0\to K^-\mu^+\nu_\mu$ are given in three bins
\begin{equation}
    \label{eq:pheno:bins}
    \text{low-$q^2$:}\,  q^2<7,\qquad
    \text{high-$q^2$:}\, q^2>7,\qquad
    \text{full:}\, \text{all $q^2$}\ ,
\end{equation}
while for the normalisation mode $B_s^0\to D_s^- \mu^+\nu_\mu$ the whole $q^2$ range is always taken. We adjust the central values and the uncertainty due to the $D_s^-\to K^+K^-\pi^-$ branching ratio according to
the most recent world average of this quantity: $\mathcal{B}(D_s^-\to K^+K^-\pi^-)=(5.37\pm 0.10)\%$~\cite{ParticleDataGroup:2022pth}.
This results in a minute shift in the central value but reduces the uncertainty by a factor of $2/3$: 
\begin{align}
    R_{\rm BF}^{\rm low} & = \left(1.65\pm 0.08({\rm stat})\pm0.07({\rm syst})\pm0.03(D_s)\right)\cdot 10^{-3} \ ,\\
    R_{\rm BF}^{\rm high} & =  \left(3.24\pm 0.21({\rm stat})\pm0.17({\rm syst})\pm0.06(D_s)\right)\cdot 10^{-3} \ .
\end{align}

To extract the ratio $|V_{ub}/V_{cb}|$, we follow Ref.~\cite{LHCb:2020ist} in defining
\begin{equation}
    \textrm{FF}_Y \equiv |V_{xb}|^{-2} \; \int\frac{\textrm{d} \Gamma(B_s^0\rightarrow Y \mu^+ \nu_\mu)}{\textrm{d}q^2}\textrm{d}q^2 \ ,
\end{equation}
where $Y=K^-,D_s^-$ and $x=u,c$, respectively. 

The posterior predictions of ${\rm FF}_{K}$
are determined in the three bins of \autoref{eq:pheno:bins}. We find
\begin{equation}
    \label{eq:FFK}
    \begin{aligned}
    \textrm{FF}_K (q^2<7 \; {\rm GeV}^2) & = 3.27 \pm 0.29\;{\rm ps}^{-1}\ , &
    \textrm{FF}_K (q^2>7\; {\rm GeV}^2)  & = 4.63 \pm 0.32 \;{\rm ps}^{-1}\ , \\
    \textrm{FF}_K ({\rm full} \; q^2 \, {\rm range})     & = 7.91 \pm 0.57\; {\rm ps}^{-1}\ , &
    \textrm{FF}_{D_s} ({\rm full} \; q^2 \, {\rm range}) & = 9.14\pm 0.35\; {\rm ps}^{-1}\ .
    \end{aligned}
\end{equation}
We note that our determination of $\textrm{FF}_{D_s}$ for the full range is consistent with the one used by the LHCb collaboration ${\rm FF}_{D_s} = 9.15\pm 0.37 \;{\rm ps}^{-1}$ \cite{LHCb:2020ist} based on the same form factors and parametrisation by the HPQCD collaboration~\cite{McLean:2019qcx}.

Finally, we can extract the ratio of the CKM elements $|V_{ub}/V_{cb}|$ using the LHCb measurements of $R_{\rm FF}$ in the different $q^2$ bins through
\begin{equation}
    \left|\frac{V_{ub}}{V_{cb}}\right| = \sqrt{R_\textrm{BF}\; \times \; R_{\rm FF} }\ ,
\end{equation}
where $R_\textrm{FF} = \textrm{FF}_{D_s}/\textrm{FF}_K$.  
The theoretical uncertainty on the CKM ratio is directly given by the spread of the predicted samples of $\sqrt{R_\textrm{FF}}$.
We obtain the total uncertainty by combining this theoretical uncertainty with the experimental uncertainty on $R_\textrm{BF}$ in quadrature via
\begin{equation}
    \sigma^2\left(\left|\frac{V_{ub}}{V_{cb}}\right|\right)
         = \sigma^2\left(\left|\frac{V_{ub}}{V_{cb}}\right|\right)_\text{th} + \sigma^2\left(\left|\frac{V_{ub}}{V_{cb}}\right|\right)_\text{exp}
\end{equation}
where
\begin{equation}
\begin{aligned}
    \sigma\left(\left|\frac{V_{ub}}{V_{cb}}\right|\right)_\text{th}
        & = \sqrt{R_{\rm BF}} \times \sigma(\sqrt{R_\textrm{FF}}) \ , \\
    \sigma\left(\left|\frac{V_{ub}}{V_{cb}}\right|\right)_\text{exp}
        & = \sqrt{R_\textrm{FF}}\times \sigma(\sqrt{R_{\textrm{BF}}})  \ .
\end{aligned}
\end{equation}
We obtain our nominal results as 
\begin{align}\label{eq:ourres}
    \left|\frac{V_{ub}}{V_{cb}}\right|_{q^2<7 \;\gev^2}^\text{LCSR+LQCD} & = 0.0681 \;\pm\; 0.0033\big|_\text{th} \pm 0.0023\big|_\text{exp} = 0.0681\pm 0.0040\\
    \left|\frac{V_{ub}}{V_{cb}}\right|_{q^2>7 \;\gev^2}^\text{LCSR+LQCD} & = 0.0801 \;\pm\; 0.0032\big|_\text{th} \pm 0.0034\big|_\text{exp} = 0.0801\pm 0.0047 \ .
\end{align}
We observe that the low-$q^2$ bin yields much smaller values for the ratio than the high-$q^2$ bin. This is consistent with the ratio from LCSR only, which yields
\begin{align}
    \left|\frac{V_{ub}}{V_{cb}}\right|_{q^2<7 \;\gev^2}^\textrm{LCSR}  = 0.057 \pm 0.005 \ , \quad\quad     \left|\frac{V_{ub}}{V_{cb}}\right|_{q^2>7\; \gev^2}^\textrm{LCSR}  = 0.068 	\pm 0.021 \ ,
\end{align}
and thus much smaller values than our combined fit. Considering only LQCD inputs, we have 
\begin{align}
    \left|\frac{V_{ub}}{V_{cb}}\right|_{q^2<7 \;\gev^2}^\textrm{LQCD}  =  0.087 \pm 0.020 \ , \quad\quad 
    \left|\frac{V_{ub}}{V_{cb}}\right|_{q^2>7 \;\gev^2}^\textrm{LQCD}  =  0.087 \pm 0.006 \ .
\end{align}
We point out that for this determination, we use both the RBC/UKQCD and HPQCD results, thereby going beyond the extraction done in Ref.~\cite{Flynn:2023qmi}. While both determinations are exactly equal, we highlight that the low-$q^2$ determination exhibits
a $3$ times larger uncertainty than the high-$q^2$ determination.

The LHCb collaboration does not provide the experimental correlation between the low and high-$q^2$ bins.
Hence, quantifying the level of agreement between their determinations of the CKM ratio cannot be done rigorously.
Disregarding any correlation, we obtain compatibility only at the $3.8\sigma$ level for the LHCb determination of the CKM ratio in \cite{LHCb:2020ist} also quoted in \autoref{eq:lhcbrat}.
Under the same caveats, our nominal determinations are compatible with each other at the $1.9\sigma$ level,
representing improved compatibility between the different bins and reducing the tension between the two determinations. In addition, our nominal high-$q^2$ result is almost a factor of 2 more precise than the LHCb result.
For the LCSR and LQCD determinations, compatibilities at the $0.5\sigma$ and $0\sigma$ level are reached.

\subsection{Comparison with other determinations}
Given the long-standing puzzles in both $|V_{cb}|$ and $|V_{ub}|$, we do not attempt to make a comprehensive comparison with all the different $V_{cb}$ and $V_{ub}$ determinations currently available. However, we make a few comments. 

First, it is interesting to directly compare our ratio of CKM elements with other determinations of this ratio. Currently, the only available measurement of this ratio is that using the baryon decays: $\Lambda_b\to p \mu^-\bar{\nu}_\mu$ (at $q^2>15$ GeV$^2$) and $\Lambda_b \to \Lambda_c \mu^- \bar{\nu}_{\mu}$ \cite{LHCb:2015eia}. Combined with the form factors from LQCD \cite{Detmold:2015aaa} this gives
\begin{equation}
     \left|\frac{V_{ub}}{V_{cb}}\right|_{q^2>15 \;\gev^2}^{\Lambda_b\to \lbrace p, \Lambda_c\rbrace \mu^-\bar\nu} = 0.080\pm 0.006 \ ,
\end{equation}
where we added the uncertainties in quadrature and updated the central value by using the PDG world average
of $\mathcal{B}(\Lambda_c^+\to p K^-\pi^+)$~\cite{ParticleDataGroup:2022pth,Belle:2013jfq,BESIII:2015bjk}.
We find excellent agreement at the $0.01\,\sigma$ level with our $\bsk$ determination in the high-$q^2$ region,
which is dominated by the LQCD form factor determinations.
For the low-$q^2$ bin our determination differs from the above by $1.7\,\sigma$.

While we recommend comparing only with determinations of ratios of the CKM elements, it is also possible to compare our results with absolute determinations of CKM elements in exclusive modes. Since we are measuring exclusive decays, with $\bsk$ as a new element, it seems most obvious to extract from the ratio a value of $|V_{ub}|$ by multiplying with a specific exclusive $|V_{cb}|$ determination. Given the current tension in the $B\to D^*$ form factors required for the latter (see e.g. Ref.~\cite{MJungCERNFlavourTH:2023} for a recent discussion), we do not include here the most recent $V_{cb}$ determinations from $D^*$. Considering only $B\to D$ exclusive transitions leads to \cite{Bigi:2016mdz} 
\begin{equation}\label{eq:Vcbex1}
    |V_{cb}|_{{\rm excl}, B\to D} = (40.49 \pm 0.97)\cdot 10^{-3} \ .
\end{equation}
This determination is compatible with the inclusive determinations \cite{Bordone:2021oof,Bernlochner:2022ucr} at less than $2\sigma$.
We obtain from \autoref{eq:ourres}
\begin{equation}
   |V_{ub}|_{q^2<7\;{\rm GeV}^2}^{B_s\to K}  = \left(2.76 \pm 0.30\right)\cdot 10^{-3} \, \quad\quad |V_{ub}|_{q^2>7\;{\rm GeV}^2}^{B_s\to K} = \left(3.24 \pm 0.33\right)\cdot 10^{-3}
\end{equation}

Comparing with a recent determination of $|V_{ub}|$ from exclusive $b\to u$ decays including LCSR and LQCD form factors \cite{Leljak:2023gna}
\begin{equation}
    |V_{ub}|_{\rm excl}= \left(3.50^{+0.13}_{-0.12}\right)\cdot 10^{-3} \ ,
\end{equation}
we find good agreement with our determination in the high-$q^2$ bin at $0.7\sigma$ and only a $2.3\sigma$ compatibility with the low-$q^2$.

\section{Conclusion}
We predict the full set of $\bsk$ form factors using updated light-cone sum rules with an on-shell kaon
at low momentum transfer $q^2$.
Specifically, we infer information on the sum rules' duality threshold parameters $s_0$ in two models. Systematic uncertainties are then accounted for by studying the threshold-model dependence as well as the renormalisation scale dependence.
Our light-cone sum rule results are slightly shifted to larger values than those previously obtained in
the literature.  
To obtain predictions in the full semileptonic $q^2$ range, we
further combine these LCSR predictions with two lattice QCD determinations at high $q^2$, which are more precise
than the sum rule results. We employ a parametrisation that respects unitarity through two novel modifications to the
well-known BGL approach. We find a consistent description of the form factors in our nominal fit,
which yields a small yet acceptable $p$-value of $6\%$.
The strong correlations between our results (across both the form factors 
and the $q^2$ points) are the main reason that an acceptable fit quality is obtained in the combination
with lattice QCD determinations.

We apply our results for the $\bsk$ form factors to the LHCb analysis of \lhcbdecay,
which measures the normalized integrated branching fraction in two $q^2$ bins.
From this measurement, we determine nominally the ratio of CKM elements
\begin{equation}
    \left|\frac{V_{ub}}{V_{cb}}\right|_{q^2<7 \;\gev^2}  = 0.0681\pm 0.0040 \ , \quad \quad \left|\frac{V_{ub}}{V_{cb}}\right|_{q^2>7\; \gev^2} = 0.0801\pm 0.0047 \ .
\end{equation}
These are mutually compatible at the $1.9\sigma$ level and exhibit significantly less tension than the determination previously obtained by LHCb.
We stress that given the current puzzle in both extractions of $|V_{ub}|$ and $|V_{cb}|$, it is not clear to which values to compare,
and we briefly discuss the implications of our new determination of their ratio.

Given the ongoing puzzle in the determination of either CKM matrix element, the tension between the two determinations
of their ratio, and the recent interest in $b\to u\ell^-\bar\nu$ processes due to their potential BSM reach,
we strongly recommend an update of the experimental analysis of this decay.
In particular, a determination of the shape of the $q^2$ distribution in \lhcbdecay decay would be
instrumental in improving our understanding of the form factors as well as in constraining potential BSM effects.

\acknowledgments

We thank Vladimir Braun and Alexander Lenz for private communications on the spectator-mass dependence
in the RGE for the twist-three kaon LCDA parameters.
We thank Méril Reboud for helpful discussions and a thorough review of our modifications to the \EOS software.
D.v.D.~acknowledges support by the UK Science and Technology Facilities Council
(grant numbers ST/V003941/1 and ST/X003167/1).

\bibliographystyle{jhep} 
\bibliography{refs.bib}

\end{document}